\documentclass{elsart}
\usepackage{natbib}
\usepackage{graphicx}
\usepackage{epsfig}

\begin{document}
\begin{frontmatter}
\title{Estimates of multipolar coefficients to search for cosmic ray 
anisotropies with non-uniform or partial sky coverage.}
\author[LPNHE]{Pierre Billoir}
\author[IPN]{\& Olivier Deligny}

\address[LPNHE]{LPNHE (Universit\'es Paris 6 \& 7, CNRS-IN2P3), Paris, France}
\address[IPN]{IPN (Universit\'e Paris Sud, CNRS-IN2P3), Orsay, France}

\begin{abstract}
We study the possibility to extract the multipolar moments of an
underlying distribution from a set of cosmic rays observed with 
non-uniform or even partial sky coverage. We show that if the 
degree is assumed to be upper bounded by $L$, 
each multipolar moment can be recovered whatever the coverage, 
but with a variance increasing exponentially with the bound $L$ if the 
coverage is zero somewhere. Despite this limitation, we show
the possibility to test predictions of a model without any assumption
on $L$ by building an estimate of the covariance matrix seen through
the exposure function.
\end{abstract}
\end{frontmatter}

\section{Introduction}

Anisotropy in the arrival directions of cosmic rays is a major
observable to understand their origin. Magnetic fields bend their 
trajectories in such a way that transport of cosmic rays is mainly 
diffusive up to high energies: this makes their angular distribution 
isotropic. Nevertheless, above the so-called knee of cosmic rays up 
to the ankle, there are predictions for small but increasing anisotropies 
with energy, predictions which of course depend on the regular and the 
turbulent components of the assumed galactic magnetic field, 
as well as the assumed distribution of sources and composition of 
cosmic rays \cite{ptuskin,candia}.
Further, at ultra-high energies, cosmic ray arrival directions
are expected to be less and less smeared out by galactic and
extragalactic magnetic fields, leading to a possible extraction 
of informations about the position of the sources 
\cite{isola,sigl,eric,dolag,demarco}. 
Hence, it is clear that any evidence for an anisotropy, 
or any limit on anisotropies in the cosmic ray locations observed 
by experiments are among the most important constraints upon models. 

The multipole expansion up to a given order $L$ is a powerful 
tool to study the structures standing out the noise down to 
an angular scale $\approx \pi/L$, whatever the shape of the underlying 
celestial pattern. In practice, the number of significant coefficients 
is limited by the angular resolution of the detector and, in the 
other hand, by the available statistics of observation. However,
ground based experiments cover a limited range in declination, so 
that it is impossible to apply off the shelf the formalism of 
multipole moments: anyone of the coefficients may be modified in 
an unpredictable way by the unseen part of the sky. Methods have 
been developped to study the CMB with an incomplete coverage 
\cite{gorski,wright,tegmark,mortlock}, 
but here we are faced to a different problem: we cannot suppose
a priori that the distribution of cosmic rays is described by a 
power spectrum, because we want to detect possible non-isotropic 
structures, a priori unknown. In other terms, the information 
carried by the $a_{\ell m}$ cannot be reduced to the only 
knowledge of the $C_{\ell}$.


One purpose of this paper is to study the possibility of estimating the 
multipole moments of a distribution of points over a sphere
in case of a non-uniform or even a partial coverage of the sky, together
with the limitations of such an approach.
The estimation of dipoles and quadrupoles was studied in 
\cite{sommers,julien,silvia}. Here, we use the moments of 
the observed distribution on a set of orthogonal
functions: either the spherical harmonics themselves, or a set of
functions tailored on the coverage function.
With these two different methods, we show that the interference
between the modes induced by the the non-uniformity or the
hole of the coverage can be removed assuming a bounded expansion
in the conjugate space, allowing to recover the underlying 
multipole moments. However, in accordance with the simple 
intuition that it is impossible to describe the unseen part of
the sky, we point out that the uncertainty
on the recovered coefficients increases with the assumed bound
$L$ of the expansion. We show that the larger the hole in the
coverage of the sky, the faster  the increase of uncertainty with
$L$. After some general considerations
about the description of point processes on a sphere in Section 2, 
Sections 3 and 4 are dedicated to these methods whereas Section 5
illustrates them with some examples.

Because of the incomplete knowledge of the distribution 
of cosmic ray sources, and the stochastic nature of the 
propagation through magnetic fields, the anisotropies we 
want to characterize are not reducible to explicit models: they
may be interpreted as a particular realization of a random process. 
This means that some model predictions are better expressed as 
average values of the coefficients, with their covariance matrix. 
This matrix is not necessarily diagonal to describe the physics 
we are interested in, contrary to the case of a power spectrum. 
We show in Sect.6 that under reasonable assumptions, an estimate 
can be performed with a partial sky coverage, evading the problem 
of setting a bound to the expansion.


\section{Generalities about point processes on a sphere}
 
The number of cosmic rays $n(\theta,\varphi)$ observed as a 
function of $\Omega=(\theta,\varphi)$ is a random process that 
we can modelize with the following quantity :\[
  n(\Omega)=\frac{1}{N}\sum_{i=1}^N\delta(\Omega,\Omega_i)
\]
where $\delta$ is the Dirac function on the surface of the unit 
sphere, and $\Omega_i$ the position of the $i^{th}$ cosmic ray. 
This distribution follows a Poisson law with an averaged
density that we will denote by $\mu(\Omega)$: \[
  \mu(\Omega)=\omega(\Omega)\lambda(\Omega).
\] 
Here, $\lambda$ is the density of the distribution of cosmic rays
and $\omega$ is the exposure function of the experiment.
The multipole coefficients of the function $\lambda(\theta,\varphi)$ 
defined on the unit sphere express its expansion in spherical 
harmonics: \[  
  \lambda(\theta,\varphi) = \sum_{\ell,m} a_{\ell m}\, 
  Y_\ell^m(\theta,\varphi)~~~~~
  (\ell \geq 0,~-\ell \leq m \leq \ell). 
\]
In this paper, we choose to normalize the spherical harmonics in
such a way that $\int\mathrm{d}\Omega Y_{\ell m}(\Omega)
Y_{\ell' m'}(\Omega) = 4\pi \delta_{\ell\ell'}\delta_{m m'}$. 
Together with the normalization $\int\mathrm{d}\Omega \lambda(\Omega)=4\pi$,
our convention leads to $a_{00}=1$ which is, in the context of
this study, a natural system of units. For convenience, we will use 
hereafter the notation $\sum_{\ell,m}=\sum_{\ell=0}^\infty\sum_{m=-\ell}^\ell$. 

With a uniform sky coverage, it is easy to obtain an unbiased evaluation of
these coefficients from a sample of $N$ points ($\theta_i,\varphi_i$)
distributed independently according to the density $\lambda$ : \[ 
  \overline{a}_{\ell m} = \frac{1}{N} \sum_{i=1}^N  Y_\ell^m(\theta_i,\varphi_i).
\]
If the distribution is roughly uniform (that is, $|a_{\ell m}| \ll 1$ 
for all $(\ell,m) \neq (0,0)$), these estimators are quasi-optimal, 
weakly correlated and their variances are close to $1/N$; otherwise the 
variances can be approximated from the quadratic moments: \[
  {\rm var}(\overline{a}_{\ell m}) = 
  \frac{1}{N}\, \sum_{i} \bigg( \big(Y_\ell^m(\theta_i,\varphi_i)\big)^2 - 
  (\overline{a}_{\ell m})^2 \bigg). 
\]

These properties are due to the orthogonality of the spherical 
harmonics, and cannot been used directly if the coverage of the sphere 
is not uniform, that is, if the distribution actually observed is 
$\lambda(\theta,\varphi)\omega(\theta,\varphi)$, where $\omega$ is a 
non-uniform function eventually vanishing in some regions.

However, if we suppose that the expansion of $\lambda$ in spherical 
harmonics is bounded to degree $L$ (at least in good approximation), 
we are going to see that it is possible to recover - within limitations 
that we will discuss in details - the multipolar coefficients even in 
case of partial sky coverage.

\begin{figure}[t]
	\centering
	\includegraphics[width=12cm]	{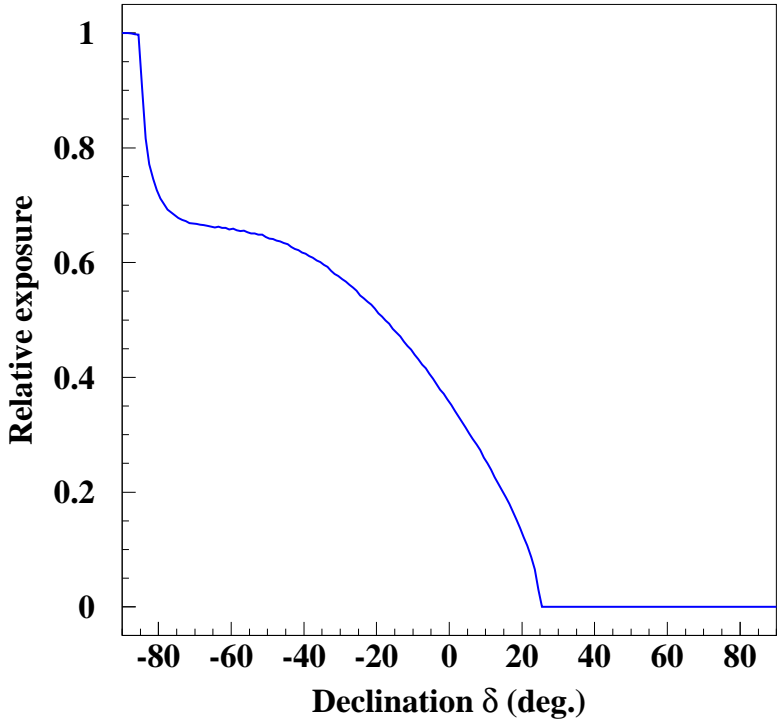}
	\caption{\small{\textit{Relative exposure as a function of declination in 
	equatorial coordinates of the Southern site of the Pierre Auger observatory. 
	The detection efficiency is assumed to be saturated up to zenithal of $60^\circ$.}}}
	\label{fig:CovAugerSud}
\end{figure}

Throughout the paper, we will consider by default an exposure function 
not covering the whole sky in a realistic way since we use the 
function calculated by Sommers \cite{sommers} describing the coverage of the sky 
of the Southern site of the Pierre Auger observatory as long as the acceptance of 
the detector is saturated until a local zenith angle $\theta_{\mathrm{max}}$.
This function is shown on Fig.\ref{fig:CovAugerSud} with 
$\theta_{\mathrm{max}}=60^\circ$, which guarantees in a realistic way this 
ideal function  to be meaningful \cite{allard}.

\section{Estimate through the deconvolution of the exposure function}
\label{Est1}

\subsection{The estimate}

In this section, we describe an estimate of the $a_{\ell m}$ 
coefficients based on the interpretation of the estimate \[
  \overline{b}_{\ell m} = \frac{1}{N}\sum_{i=1}^N  Y_\ell^m(\theta_i,\varphi_i)
  \]
in terms of a convolution between the underlying $a_{\ell m}$ 
coefficients of the density $\lambda(\theta,\varphi)$ and a kernel 
which depends on the $\omega(\theta,\varphi)$ function. In some extent,
this approach is the equivalent of the MASTER one within the CMB
framework \cite{hivon}, except that we are interested here in 
building a linear estimate of the $a_{\ell m}$ coefficients rather
than a quadratic estimate of the $C_\ell$ ones. As the cosmic 
rays are observed through the exposure function $\omega$, the estimate 
$b_{\ell m}$ is not an estimate of the multipolar coefficients of the 
density $\lambda$, but an estimate of the multipolar coefficients of
$\omega\lambda$. The $a_{\ell m}$ coefficients are thus related to the 
$b_{\ell m}$ ones through the following convolution \[
  b_{\ell m} = \sum_{\ell^{\prime},m^{\prime}} 
  [K]_{\ell \ell^{\prime}}^{mm^{\prime}}a_{\ell^{\prime}m^{\prime}}.
\]
The kernel $K$ is entirely determined by the specific exposure function. 
Indeed, by using the completeness relation of the spherical harmonics, 
the elements of the kernel $[K]_{\ell \ell^{\prime}}^{mm^{\prime}}$ read \[
  [K]_{\ell \ell^{\prime}}^{mm^{\prime}} = \int_{4\pi}\mathrm{d}\Omega~ 
  Y_\ell^m(\Omega)\omega(\Omega)Y_{\ell^{\prime}}^{m^{\prime}}(\Omega).
\]
This relation was refered to as the \textit{convolution theorem} 
in~\cite{peebles}, as this is the analog on the sphere of the convolution 
theorem for a Fourier's transform. Then, by using direct numerical results 
of $K$ and $K^{-1}$ for specific exposure function $\omega$, the underlying
$a_{\ell}^m$ coefficients can be formally recovered through the following 
estimate \[
  \overline{a}_{\ell m} = \sum_{\ell^{\prime},m^{\prime}} 
  [K^{-1}]_{\ell \ell^{\prime}}^{mm^{\prime}}\overline{b}_{\ell^{\prime}m^{\prime}}.
\]

\subsection{Statistical properties of the estimate}
\label{StatProp}

The observed $N$ points are sampled according to a Poissonian process 
on the sphere. Averaged over a large number of realisations of $N$ events 
distributed according to $\mu(\theta,\varphi)$, it's elementar to compute 
the first and the second moment of $n(\Omega)$ :
\begin{eqnarray*}
	\left\langle n(\Omega) \right\rangle_P &=&
	\mu(\Omega) \\
	\left\langle n(\Omega)n(\Omega^{\prime}) \right\rangle_P &=&
	\mu(\Omega)\mu(\Omega^{\prime}) +
	\mu(\Omega)\delta(\Omega,\Omega^{\prime})
\end{eqnarray*}
where the subscript $P$ stands for \textit{Poisson}. The average of the $b_{\ell m}$ 
estimate then reads 
\begin{eqnarray*}
	\left\langle \overline{b}_{\ell m} \right\rangle_P &=& \left\langle
	\int_{4\pi}\mathrm{d}\Omega~ n(\Omega) Y_\ell^m(\Omega) \right\rangle_P\\
	&=& \int_{4\pi}\mathrm{d}\Omega~ \mu(\Omega) Y_\ell^m(\Omega) \\
	&=& \sum_{\ell^{\prime},m^{\prime}} K_{\ell
	\ell^{\prime}}^{mm^{\prime}}a_{\ell^{\prime} m^{\prime}}
\end{eqnarray*}
leading to the following averaged $a_{\ell m}$ estimate
\begin{eqnarray*}
	\left\langle \overline{a}_{\ell m} \right\rangle_P &=& 
	\sum_{\ell_1,m_1} [K^{-1}]_{\ell\ell_1}^{mm_1}\sum_{\ell_2,m_2} 
	K_{\ell_1\ell_2}^{m_1m_2}a_{\ell_2 m_2}\\
	&=&\sum_{\ell_2,m_2} \sum_{\ell_1,m_1} [K^{-1}]_{\ell\ell_1}^{mm_1} 
	K_{\ell_1\ell_2}^{m_1m_2}a_{\ell_2 m_2}\\
	&=& a_{\ell m}.
\end{eqnarray*}
Thus, it is clear that we have built an unbiased estimate.
Turning to the covariance, we get in the same way	
\begin{eqnarray*}
	\mathrm{cov}(\overline{b}_{\ell m},\overline{b}_{\ell^{\prime} m^{\prime}})=&&
	\int_{4\pi} \mathrm{d}\Omega\mathrm{d}\Omega^{\prime}
	\mu(\Omega)\mu(\Omega^{\prime})Y_\ell^m(\Omega)
	Y_{\ell^{\prime}}^{m^{\prime}}(\Omega^{\prime})\\
	&+&\int_{4\pi} \mathrm{d}\Omega\mathrm{d}\Omega^{\prime}
	\mu(\Omega)\delta(\Omega,\Omega^{\prime})Y_\ell^m(\Omega)
	Y_{\ell^{\prime}}^{m^{\prime}}(\Omega^{\prime})
	-\left\langle \overline{b}_{\ell m} \right\rangle_P 
	\left\langle \overline{b}_{\ell^{\prime} m^{\prime}} \right\rangle_P. 
\end{eqnarray*}
The only non vanishing term comes from the Poissonian part of the second
moment of $n(\Omega)$:
\begin{eqnarray*}
	\mathrm{cov}(\overline{b}_{\ell m},\overline{b}_{\ell^{\prime} m^{\prime}}) =
	\sum_{\ell_1,m_1} a_{\ell_1m_1} \int_{4\pi}\mathrm{d}\Omega~
	Y_\ell^m(\Omega)Y_{\ell^{\prime}}^{m^{\prime}}(\Omega)
	\omega(\Omega)Y_{\ell_1}^{m_1}(\Omega).	
\end{eqnarray*}
Using the fact that we are in practice looking for small deviation per respect
with isotropy as emphasized in the introduction (ie: $a_{\ell m}/a_{00} \ll 1$), 
this expression can be simplified to:\[
  \mathrm{cov}(\overline{b}_{\ell m},\overline{b}_{\ell^{\prime} m^{\prime}}) =
  [K]_{\ell \ell^{\prime}}^{mm^{\prime}}a_{00} ,
\]
leading to :\[
  \mathrm{var}(\overline{a}_{\ell m}) =
  [K^{-1}]_{\ell \ell}^{mm}a_{00} .
\]
Let's remind that $K$ being proportional to the number of events, the 
standard deviation of the reconstructed coefficients is hence proportional to 
$1/\sqrt{N}$ as expected.
In case of a non-uniform \textit{but full coverage} of the sky, the completeness 
relation of the spherical harmonics easily allows to give the following analytical expression of the $K^{-1}$ operator : \[
  [K^{-1}]_{\ell \ell'}^{mm'} = \int_{4\pi}\mathrm{d}\Omega \frac{1}{\omega(\Omega)}
  Y_\ell^m(\Omega)Y_{\ell^{\prime}}^{m^{\prime}}(\Omega).
\]
In case of partial coverage, the spherical harmonics are no longer orthogonal, in
such a way that the coefficients of $K^{-1}$ only satisfy the expression  \[
  \sum_{\ell_1,m_1}^L [O]_{\ell\ell_1}^{mm_1}[K^{-1}]_{\ell_1\ell'}^{m_1m'}
  = \int_{\Delta\Omega}\mathrm{d}\Omega \frac{1}{\omega(\Omega)} 
  Y_\ell^m(\Omega)Y_{\ell^{\prime}}^{m^{\prime}}(\Omega)	
\]
where $\Delta\Omega$ is the non-zero region of $\omega$, and \[
  [O]_{\ell\ell'}^{mm'}=\int_{\Delta\Omega}\mathrm{d}\Omega
  Y_\ell^m(\Omega)Y_{\ell'}^{m'}(\Omega).
\] 
It is then obvious, in this latter case, that $K^{-1}$ is invertible 
only if $L$ is \emph{finite}, and that the coefficients of $K^{-1}$ strongly 
depend on the assumed bound $L$, leading to an indetermination of each 
coefficient as $L$ is increasing. This indetermination is nothing else but 
the mathematical traduction that it's impossible to know the distribution 
of cosmic rays in the uncovered region of the sky.

\section{Estimate through dedicated orthogonal functions}
\label{Est2}

In this section, we describe another way, more intuitive, to recover 
the underlying $a_{\ell m}$ coefficients by applying the Gram-Schmidt 
procedure to the $\omega(\theta,\varphi)Y_\ell^m(\theta,\varphi)$ 
with $\ell \leq L$, which allows to build orthogonal functions from the 
coverage function. Then, by applying the formalism of moments 
to these functions; the $a_{\ell m}$ are obtained with linear combinations 
of these moments.

\subsection{Applying the Gram-Schmidt procedure}

The scalar product being defined as \[
  \langle f | g \rangle =
  \frac{1}{4\pi} \int f^*(\theta,\varphi)\, g(\theta,\varphi)\,d\Omega, 
\]
the normalized spherical harmonics may be written as \[
  Y_\ell^m(\theta,\varphi) = P_\ell^m(\cos\theta)\,e^{im\varphi} 
\]
where the $P_\ell^m$ are the associated Legendre functions supposed here to be
normalized: \[
\frac{1}{2} \int_{-1}^{1}P_\ell^m(x)^2\,dx = 1. 
\]
In practical computations we use the real functions $Y_\ell^0(\theta,\varphi)$
for $m=0$, and $\{\sqrt{2}\,P_\ell^m(\theta,\varphi)\,\cos(m\varphi),
\sqrt{2}\,P_\ell^m(\theta,\varphi)\,\sin(m\varphi)\}$ for $1 \leq m \leq \ell$.
For convenience we keep the notations with the $Y_\ell^m$ hereafter.

Let us suppose first that $\omega$ is a function of $\theta$ only 
(for example, if the coverage is uniform in right ascension).
Then $\omega Y_\ell^m$ and $\omega Y_{\ell'}^{m'}$ are orthogonal if 
$m \neq m'$, and the orthogonalisation may be performed separately 
for each value of $m$, combining the $\omega Y_\ell^m$ with 
$m \leq \ell \leq L$. If $\mathcal{N}(f)$ represents the
function $f$ after normalization, we just need to set, for a given $m$:
\begin{eqnarray*}
Q_{|m|}^m & = & \mathcal{N}(\omega P_{|m|}^m) \\
Q_{{|m|}+1}^m & = & \mathcal{N}\big(\omega P_{|m|+1}^m - 
  \langle Q_{|m|}^m |\omega P_{|m|+1}^m \rangle\, Q_{|m|}^m \big) \\
Q_{|m|+2}^m & = & \mathcal{N}\big(\omega P_{|m|+2}^m - 
  \langle Q_{|m|}^m |\omega P_{|m|+2}^m \rangle\, Q_{|m|}^m 
  -\langle Q_{|m|+1}^m |\omega P_{|m|+2}^m \rangle\, Q_{|m|+1}^m \big) \\
\cdots \\
Q_{{|m|}+p}^m & = & \mathcal{N}\big(\omega P_{|m|+p}^m - 
  \sum_{k=0}^{p-1}\langle Q_{|m|+k}^m |\omega P_{|m|+p}^m \rangle\, Q_{|m|+k}^m \big)\\
\cdots
\end{eqnarray*}
Then the normalized functions $Z_\ell^m$ defined on the sphere by: \[     
  Z_\ell^m(\theta,\varphi) = Q_\ell^m(\cos\theta)\,e^{im\varphi} 
\]
are orthogonal to each other, and the subset of $Z_\ell^m$ with
$|m| \leq \ell \leq L$ generates the same subspace as the 
$\omega Y_\ell^m$ with $|m| \leq \ell \leq L$.
We can express them through a set of coefficients $C_{\ell \ell^{\prime}}^m$: \[
  Z_\ell^m(\theta,\varphi) = 
  \sum_{\ell'=m}^{\ell}C_{\ell \ell^{\prime}}^m\,\omega(\theta)
  \,Y_{\ell'}^m(\theta,\varphi). 
\]
If $\omega$ depends on both $\theta$ and $\varphi$, the same procedure can be
applied, but the orthogonal functions are mixtures of different values of $m$,
and there is no canonical way to obtain them; anyway it is possible to build
a basis preserving the subset generated by $0 \leq \ell \leq L$ whatever $L$.
For simplicity, we do not develop such a formalism here. In particular, as 
only a small dependence on $\varphi$ is expected in the case we are interested in, it
is possible to weight the events to account for this variation of the exposure as a
function of the right ascension, and hence, the formalism applied here can be applied
off the shelf.

\begin{figure}[t]
	\centering
	\includegraphics[width=14cm]{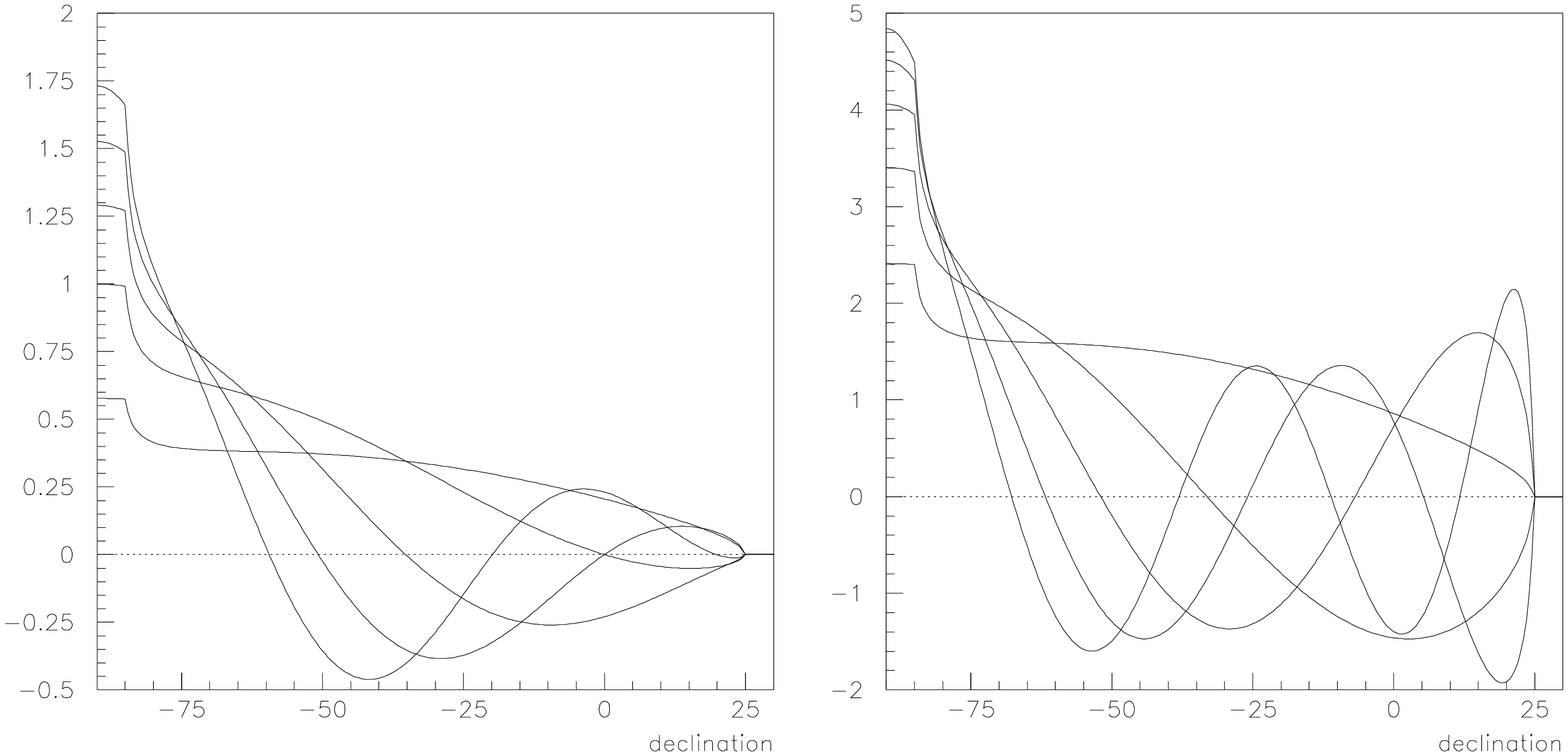}
	\includegraphics[width=6.5cm]{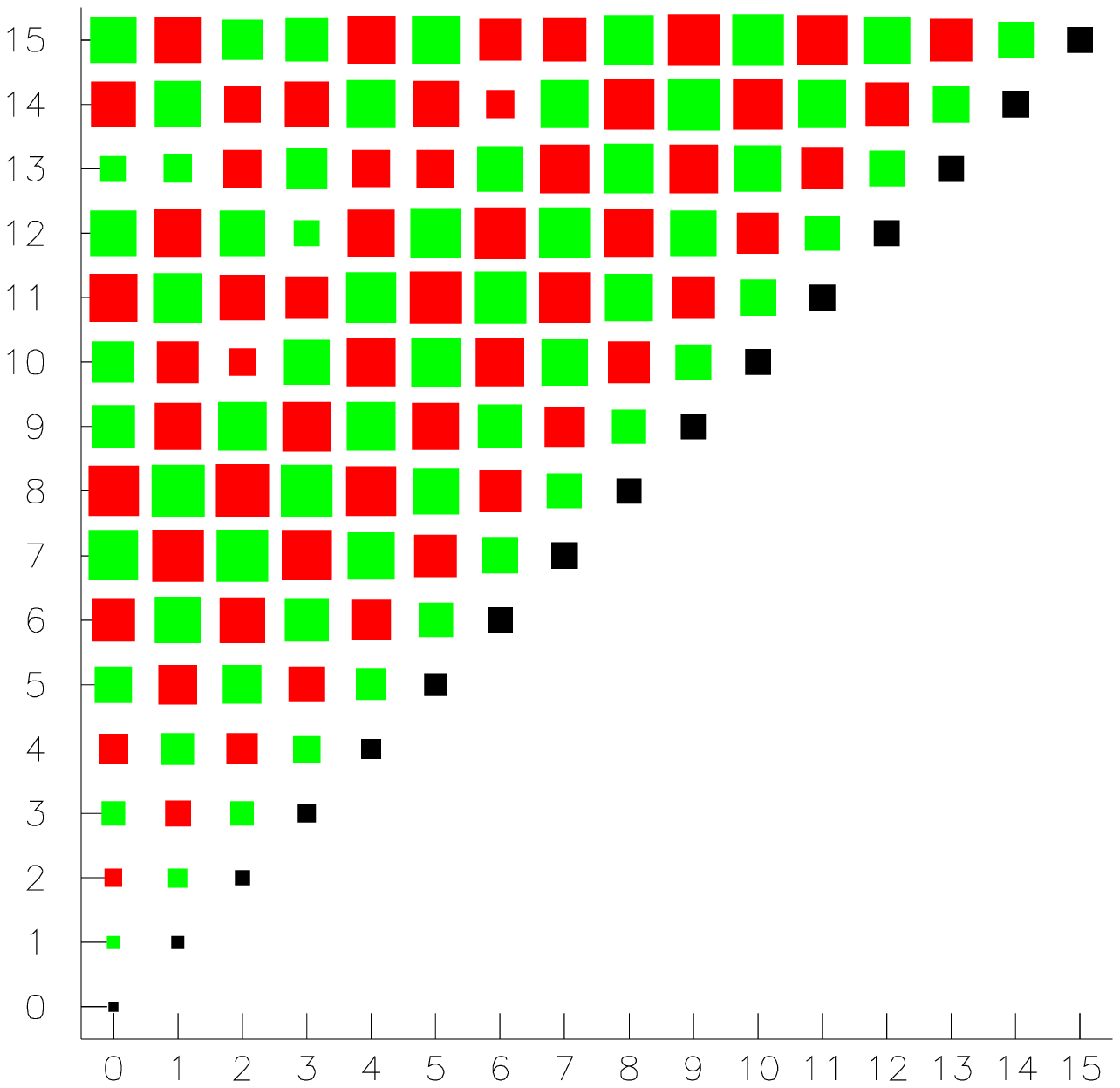}
	\includegraphics[width=6.5cm]{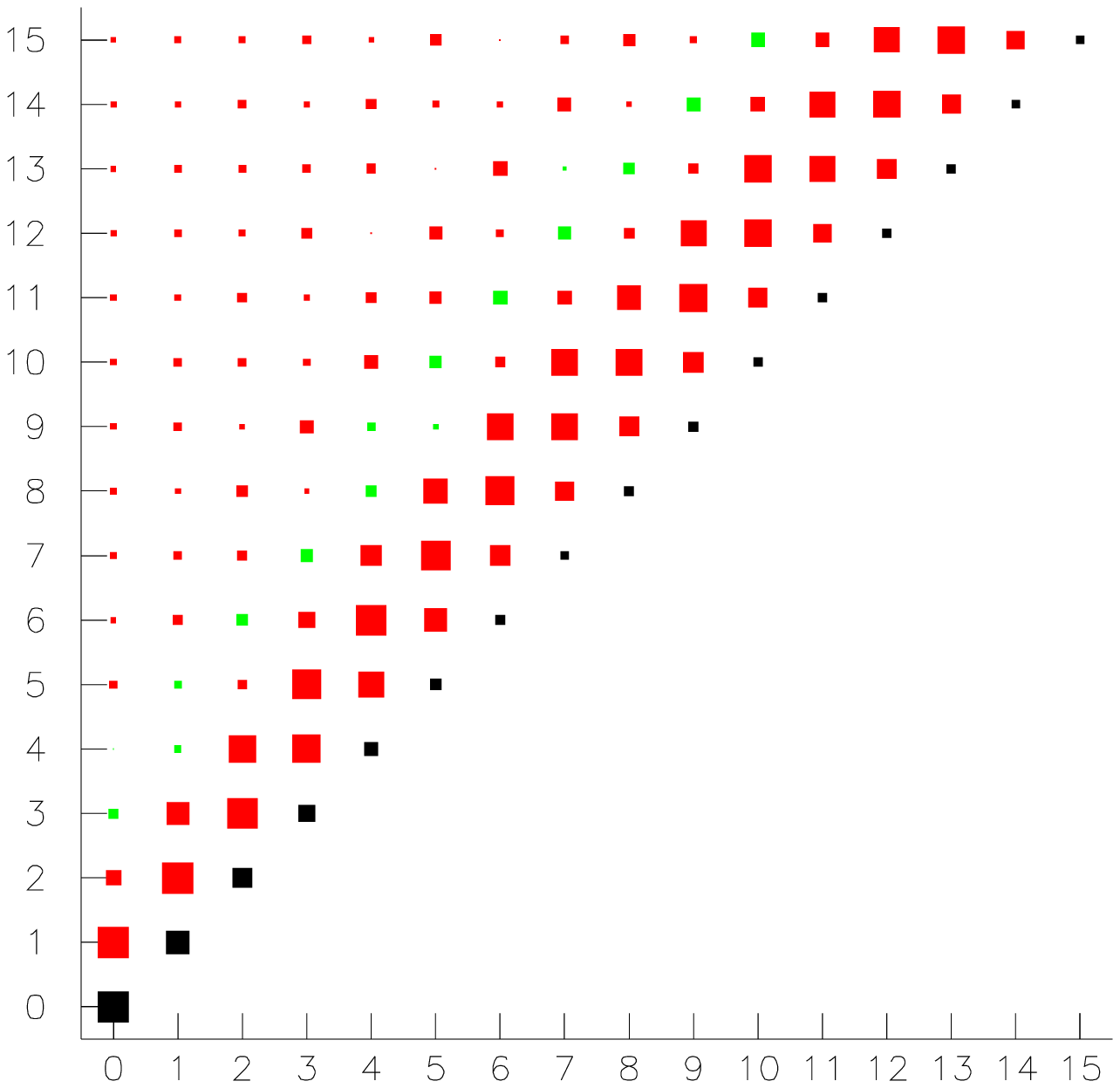}
	\caption{\small{\textit{Transformation between the $\omega Y_\ell^0$ 
	and the $Z_\ell^0$ with the coverage function of
	Fig.\ref{fig:CovAugerSud}. Top: shape (as a function of the
        declination) for $\ell \leq 4$ ($\omega Y_\ell^0$ on left side,
        $Z_\ell^0$ on right side). Bottom: coefficients  for $\ell \leq 15$;
        left: the $C_{\ell \ell^{\prime}}^0$, in logarithmic 
	scale: the area (in the units of the axes) is $\ln|C_{\ell \ell^{\prime}}^0|/10$ 
	(that is: a point represents 1, a unit square represents $2.2\times 10^4$); the 
	off-diagonal coefficients are in green if negative, in red if positive; right: the 
	$D_{\ell \ell^{\prime}}^0$ (inverse matrix) in linear scale 1:1, with the same 
	sign convention.}}}
	\label{fig:coeff_llm}
\end{figure}

To illustrate the method, Fig.\ref{fig:coeff_llm} displays the
shape of the $\omega Y_\ell^m$ and the $Z_\ell^m$, and the triangular matrix of transformation
$C_{\ell \ell^{\prime}}^m$ for $L=15,m=0$ (in logarithmic scale),
in the case of the coverage function displayed in Fig.\ref{fig:CovAugerSud}. One can
see that the off-diagonal terms (in absolute value) grow
rapidly well above 1 with $\ell$ and dominate over the diagonal ones (the coefficients
for other values of $m$ have a quite simular pattern). This strong ``mismatch'' between
the $\omega Y_\ell^m$ and the $Z_\ell^m$ is suggested by the shapes of the functions and
may be understood qualitatively in the following
way: for large values of $\ell-|m|$, the $\theta$ dependence of $Y_\ell^m$
has $\ell-|m|$ oscillations over the full interval $[0,\pi]$, while $Z_\ell^m$ has
the same number of oscillations over the covered interval; this makes difficult a
matching of $Z_\ell^m$ to functions like $\omega Y_\ell^m$ which have less oscillations over
this interval.
 
\subsection{Estimating the multipole coefficients}

Points being distributed according to the density $\lambda(\theta,\varphi)$
(to be evaluated), and detected with a probability $\omega(\theta)$ (supposed to be
known), the observed points are distributed according to $\omega\lambda$: 
this function may be expanded over the $Z_\ell^m$ defined from $\omega$ 
as explained above: \[
  \omega\lambda = \sum_{\ell,m}\,\alpha_{\ell m} Z_\ell^m 
\]
and an unbiased estimator of the $\alpha_{\ell m}$ is obtained from the points: \[
  \overline{\alpha}_{\ell m} = \frac{1}{N} \sum_{i} Z_\ell^m(\theta_i,\varphi_i). 
\]
If $\lambda$ is quasi-uniform, $\omega\lambda$ is almost proportional 
to $Z_0^0$: the coefficient $\alpha_{00}$ is largely dominant. 
Then these estimators are quasi-optimal; if $\omega$ 
is not constant, the $\alpha_{\ell m}$, for a given value of
$m$, may be correlated. If $N$ is large, their covariance matrix is 
approximately given by quadratic moments: \[
  {\rm cov}(\overline{\alpha}_{\ell m},\overline{\alpha}_{\ell^{\prime}m}) \simeq 
  \frac{1}{N}\, \sum_{i}  Z_\ell^m(\theta_i,\varphi_i)\,
  Z_{\ell^{\prime}}^m(\theta_i,\varphi_i)
  - \overline{\alpha}_{\ell m}\,\overline{\alpha}_{\ell^{\prime}m}.
\]
It is now easy to obtain estimators of the multipole coefficients of
$\lambda$ at a given order $L$ by substituting the expressions of 
the $Z_\ell^m$: \[  
  \omega\lambda \simeq \sum_{\ell,m}^L\,\overline{\alpha}_{\ell m}
  \sum_{\ell'=m}^{\ell}C_{\ell \ell^{\prime}}^m\,\omega\,Y_{\ell'}^m,
\]
that is: \[
  \lambda \simeq \sum_{\ell,m}^L\overline{a}_{\ell m}\,Y_{\ell}^m~~~~{\rm with}~~~~
  \overline{a}_{\ell m} =  \sum_{\ell'=l}^L\,C_{\ell^{\prime} \ell}^m\,      
  \overline{\alpha}_{\ell^{\prime} m}.
\]
The $\overline{a}_{\ell m}$ with different values of $m$ are not correlated, and
the covariance matrix of the $\overline{a}_{\ell m}$ is given by: \[
  {\rm cov}(\overline{a}_{\ell_1m},\overline{a}_{\ell_2m}) = 
  \sum_{\ell',\ell''}\,C_{\ell'\ell_1}^m\,C_{\ell''\ell_2}^m \,
  {\rm cov}(\overline{\alpha}_{\ell'm},\overline{\alpha}_{\ell''m}).
\]

\section{Illustrations}
\label{Illus}

\begin{figure}[t]
	\centering
	\includegraphics[width=10cm]{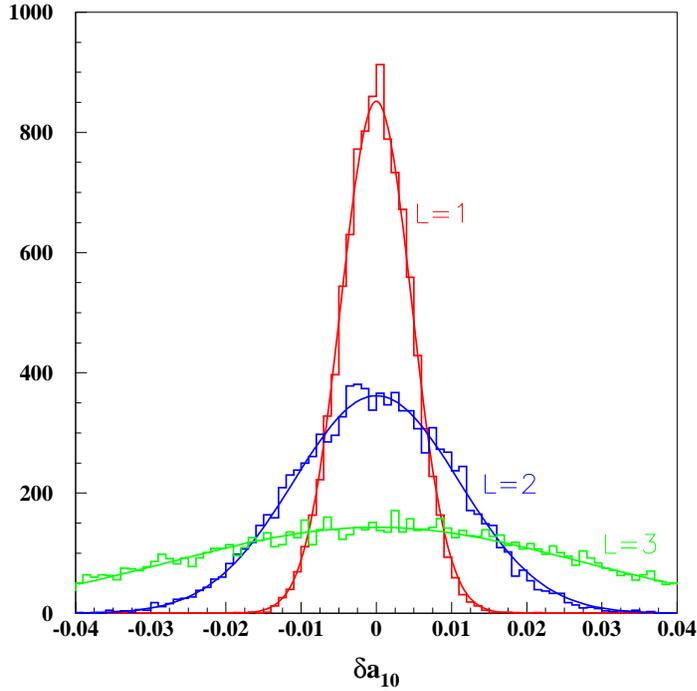}
	\caption{\small{\textit{Reconstruction accuracy of the $\overline{a}_{10}$ 
	coefficient in case of a $a_{10}=0.1$ dipolar pattern injected as a function of the
	assumed bound $L$=1 (red),2 (blue),3 (green), in case of exposure function shown on
	Fig.\ref{fig:CovAugerSud}. Histograms are from Monte-Carlo, and superimposed curves
	are Gaussian with averages and standard deviations from analytical predictions.}}}
	\label{fig:RecDipVsN}
\end{figure}

\begin{figure}[t]
	\centering
	\includegraphics[width=10cm]{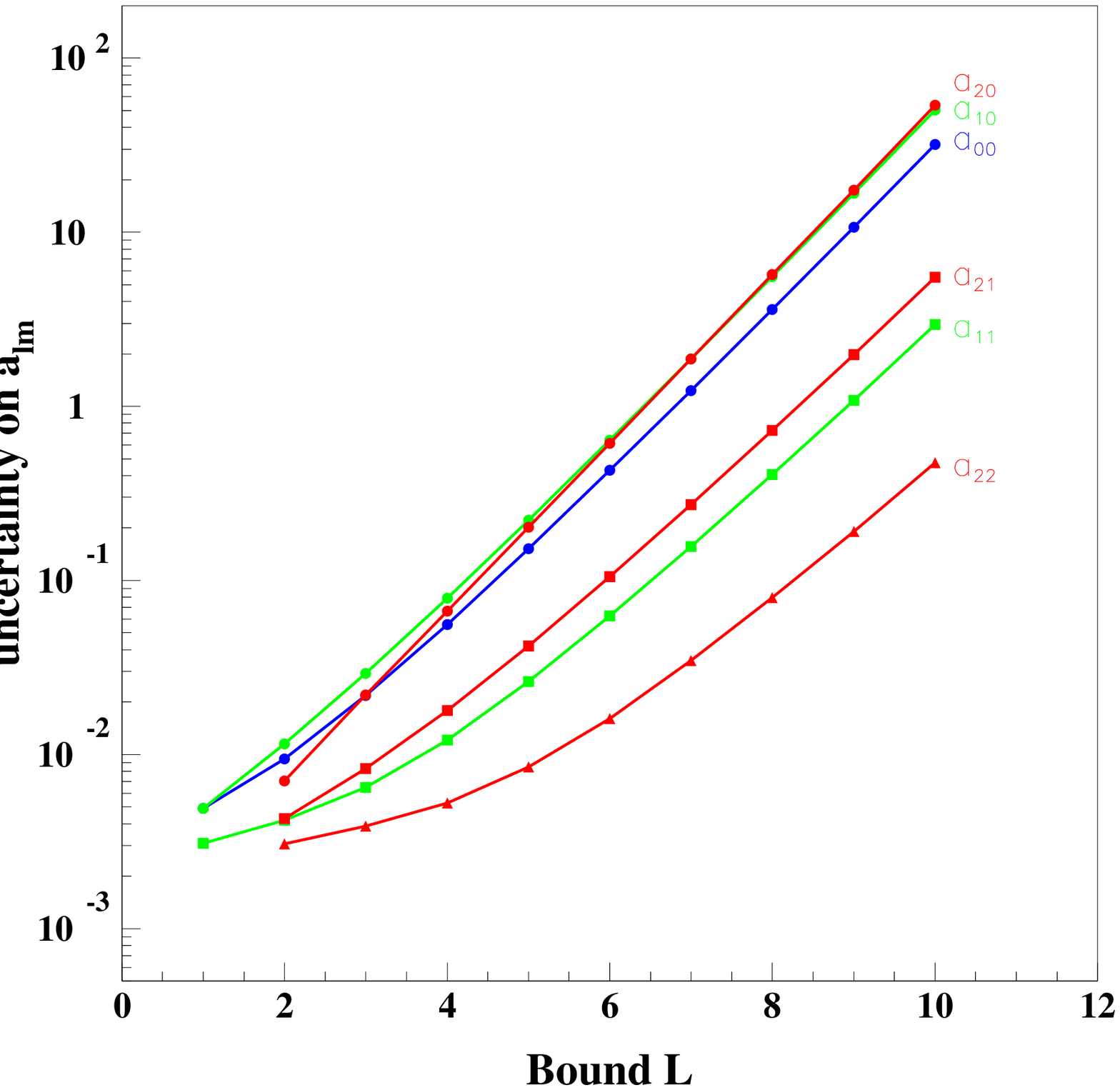}
	\caption{\small{\textit{Reconstruction accuracy of $\{a_{\ell m}\}_{\ell\leq 2}$
	 coefficients  as a function of the assumed bound $L$, in case of exposure 
	 function shown on Fig.\ref{fig:CovAugerSud}. Increasing $L$ leads to an 
	 explosion of the uncertainty on $a_{\ell m}$.}}}
	\label{fig:RecVsL}
\end{figure}

To illustrate the statistical properties of the estimates, we show here some 
simple applications of the methods in case of exposure shown on Fig.\ref{fig:CovAugerSud}.
For the sake of clarity, we will refer to as \textit{method 1} the method 
presented in section \ref{Est1}, and to as \textit{method 2} the method
presented in section \ref{Est2}.

\subsection{Behaviour of variances with $L$}

For illustrations, we use here the \textit{method 1}.
In a first time, we restrict the bound $L$ to 1, so that we are interested here
in research of a dipolar component only. We show on Fig.\ref{fig:RecDipVsN} 
the reconstruction of the coefficient $a_{10}$ in the case of an indeed 
dipolar distribution,  whose excess of events points towards equatorial North with a
magnitude $a_{10}=0.1$. The red histogram drawn show the occurence number of 
each reconstructed value of $\overline{a}_{10}$ in case of $N=10^5$ events generated 
by Monte-Carlo according to \[
  \mu(\theta,\varphi)=\omega(\theta,\varphi)[1+a_{10}Y_1^0(\theta,\varphi)].
\]
Over the histogram is plotted a Gaussian curve whose average and standard
deviation parameters are the ones determined in section \ref{StatProp}. This curve
matches the histogram, in such a way that the statistics previously determined 
by calculation describe the properties of the estimators indeed. Let us note that 
under the assumption of a purely dipolar distribution (ie $L$=1) the reconstruction 
of the multipolar coefficients is obtained in a very reasonable way. 

Let us continue to illustrate the method by looking at the same multipolar
coefficients, still in the case of a purely dipolar distribution, but by
increasing the bound $L$ to 2 and 3. Still on Fig.\ref{fig:RecDipVsN}, the blue
and the green histograms and Gaussian curves plot the same quantities than the 
red ones but for $L=2$ and $L=3$ respectively, and illustrate the extremely 
fast degradation of the accuracy of the reconstruction of $a_{10}$ by more 
than a factor 2 for each additional order.
 
This tendency to the widening of the laws is largely confirmed when one looks at 
the reconstruction of any coefficients $a_{\ell m}$ as a function
of $L$. We show this property on Fig.\ref{fig:RecVsL} for the 
$\{a_{\ell m}\}_{\ell \leq 2}$ set of coefficients, which illustrates clearly 
that it is increasingly difficult to give a meaning to the reconstructed 
values of the coefficients as soon as the maximum order of development is
greater than 3.

\begin{figure}[t]
	\centering
	\includegraphics[width=14cm,height=13cm]{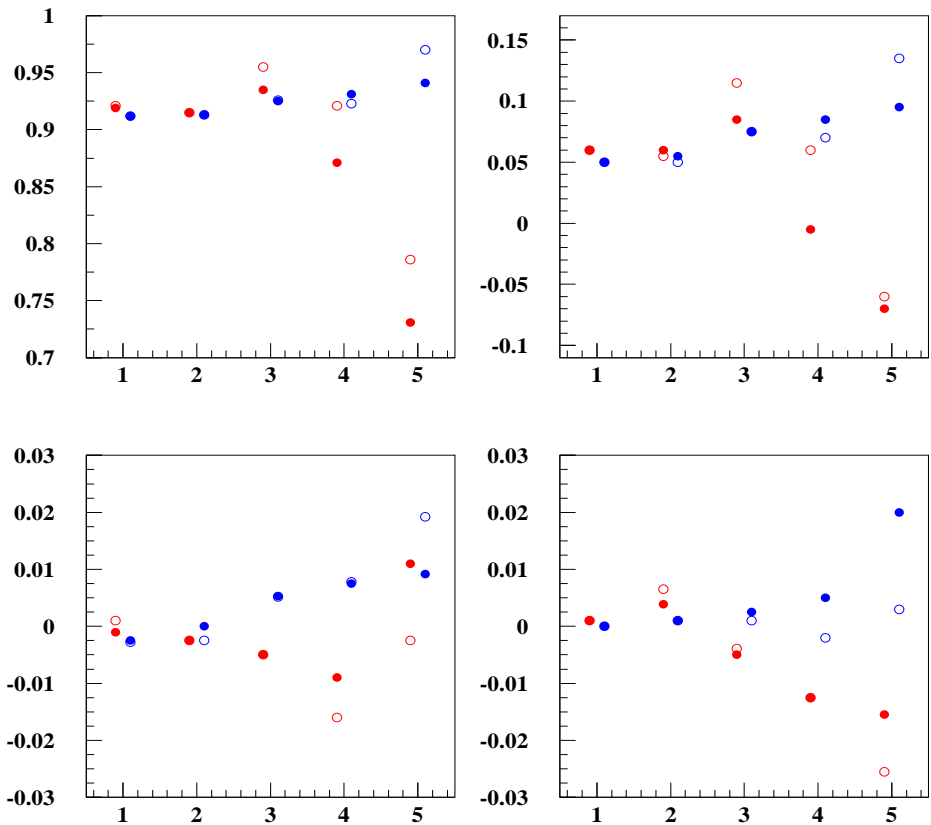}
	\caption{\small{\textit{Comparison of reconstructed coefficients $a_{\ell m}$
	($\ell=0,1$) on two simulated dipolar samples (red, blue) with the two methods
	(solid circles: method 1 presented in section \ref{Est1}; open circles: 
	method 2 presented in section \ref{Est2}), with a bound $L$ from 1 to 5. 
	Top left: $a_{00}$; top right: $a_{10}$; bottom: $a_{11}$ and $a_{1-1}$.}}}
	\label{fig:compar}
\end{figure}

\subsection{Comparison of the two methods}

Two samples of points were simulated according to a slightly anisotropic
distribution ($a_{10}/a_{00}=0.05,~a_{1\pm 1}=0$, i.e. dipole moment along $z$),
multiplied by the coverage function drawn in Fig.\,\ref{fig:CovAugerSud};
the $a_{\ell m}$ were estimated by both methods with the bound $L$ going from 1 to
5. Fig.\,\ref{fig:compar} shows that they give comparable results, and that the
difference between them is generally smaller that the intrinsic difference between
the samples (statistical fluctuations). Once again, one can see the divergence of 
the variances with increasing $L$.

\subsection{Highly non-uniform coverarge of the whole sky}

\begin{figure}[t]
	\centering
	\includegraphics[width=10cm]{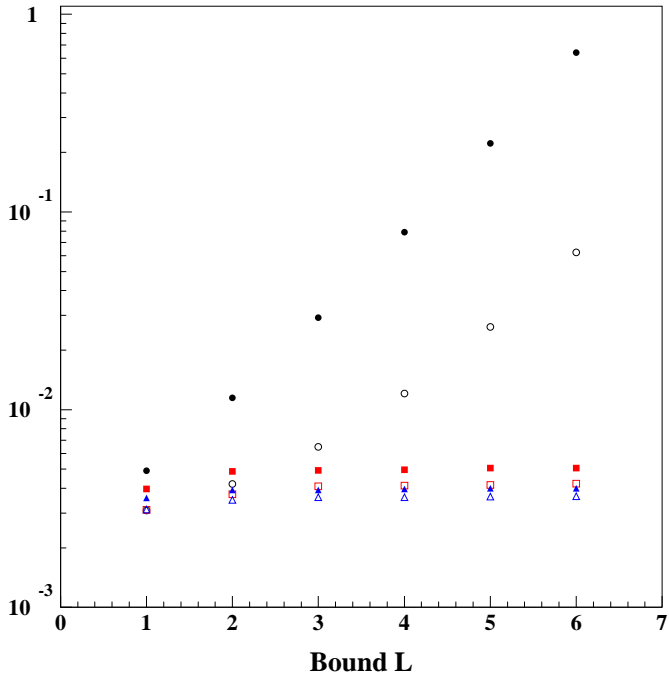}
	\caption{\small{\textit{Dependence on the bound $L$ of the variances of 
	estimated $a_{\ell m}$ with a partial or a complete (but not uniform) coverage.
	Black circles: coverage function of Fig.\,\ref{fig:CovAugerSud}; red squares: 
	the same function plus 0.1 times the symmetric one w.r.t. the equatorial plane; 
	blue triangles: the same + 0.2 times the symmetric one. The solid symbols 
	correspond to a simulated dipolar distribution along $z$ axis; the open 
	ones correspond to a axis in the equatorial plane}}}
	\label{fig:stabil}
\end{figure}

A contrario, with a complete coverage (even highly non-uniform), the size of the
variance is stabilized at large $L$. This is illustrated in Fig.\,\ref{fig:stabil},
comparing a partial coverage (cf Fig.\,\ref{fig:CovAugerSud}) to this coverage
completed by a small fraction of the same function in the opposite hemisphere,
in such a way that there is no fully unseen region. Even a relatively small relative 
exposure in the Northern part of the sky allows to recover the coefficients with
almost the same precision as if the exposure was uniform on the whole sky.
Note however that if the exposure in the opposite hemisphere tends to zero,
even if the phenomenon of stabilization at large $L$ remains, 
the variance at any $\ell$ increases, tending towards a plateau determined
roughly by $1/\sqrt{N'}$ where $N'$ is in that case the total number of 
events which would be observed on the full sky through a uniform window 
but with a low absolute coverage, in such a way that $N'$ is small. 
Of course, the larger the size of the relative exposure tending to 0 is, 
the faster the increase of the variance towards this plateau occurs.

\subsection{Angular distribution in the covered region}

\begin{figure}[t]
	\includegraphics[width=4.5cm,angle=90]{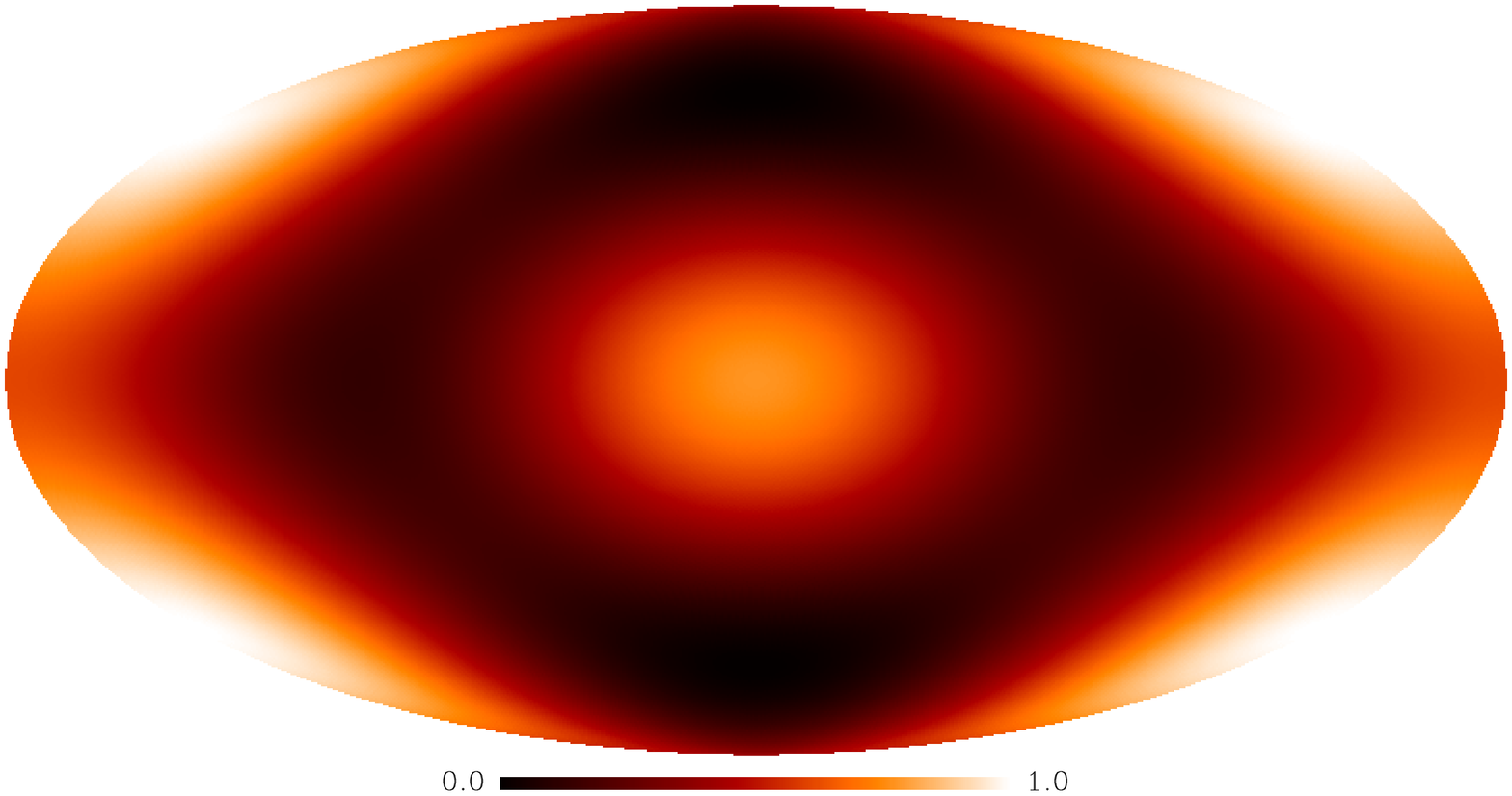}
	\includegraphics[width=4.5cm,angle=90]{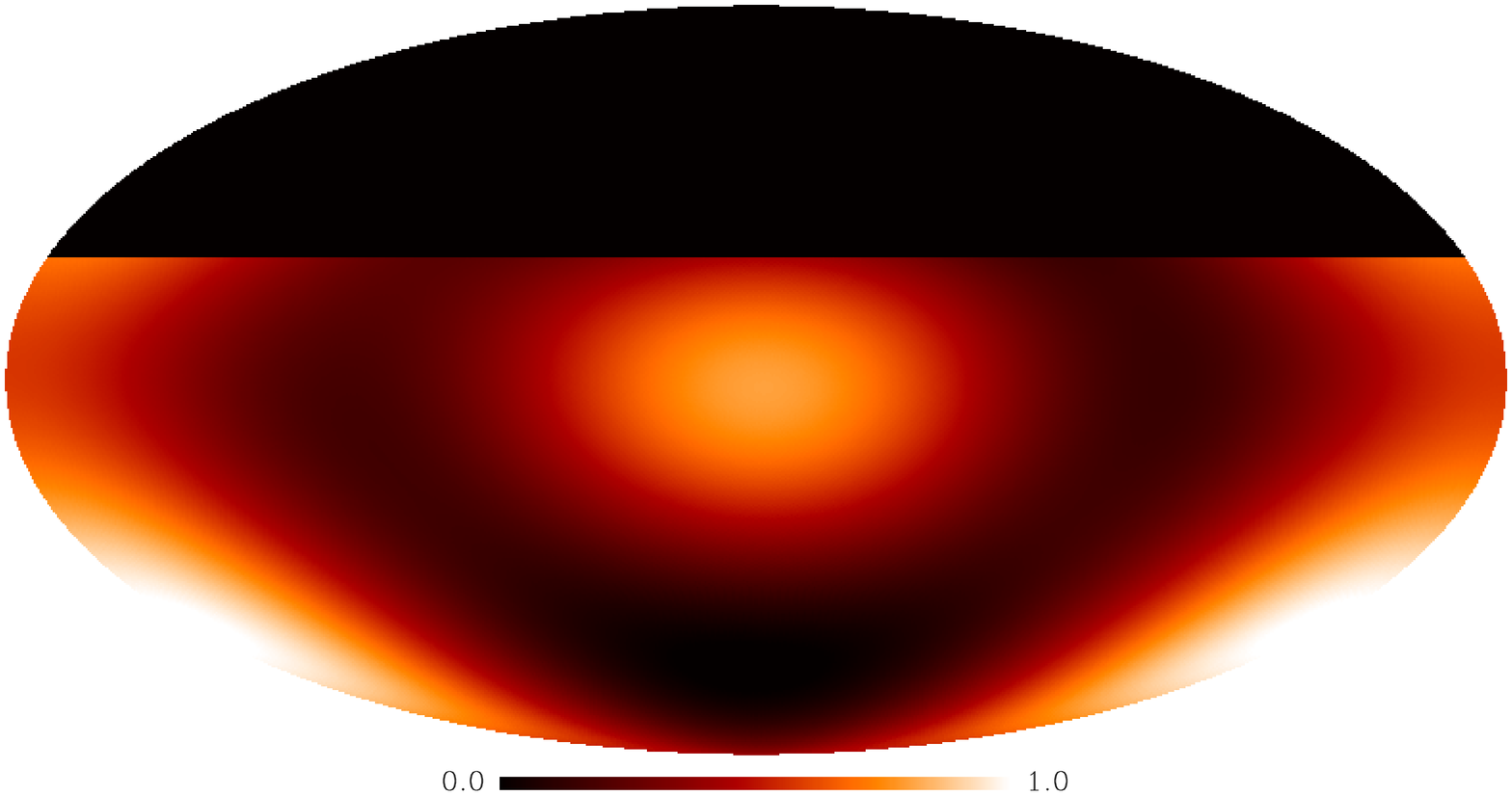}
	\includegraphics[width=4.5cm,angle=90]{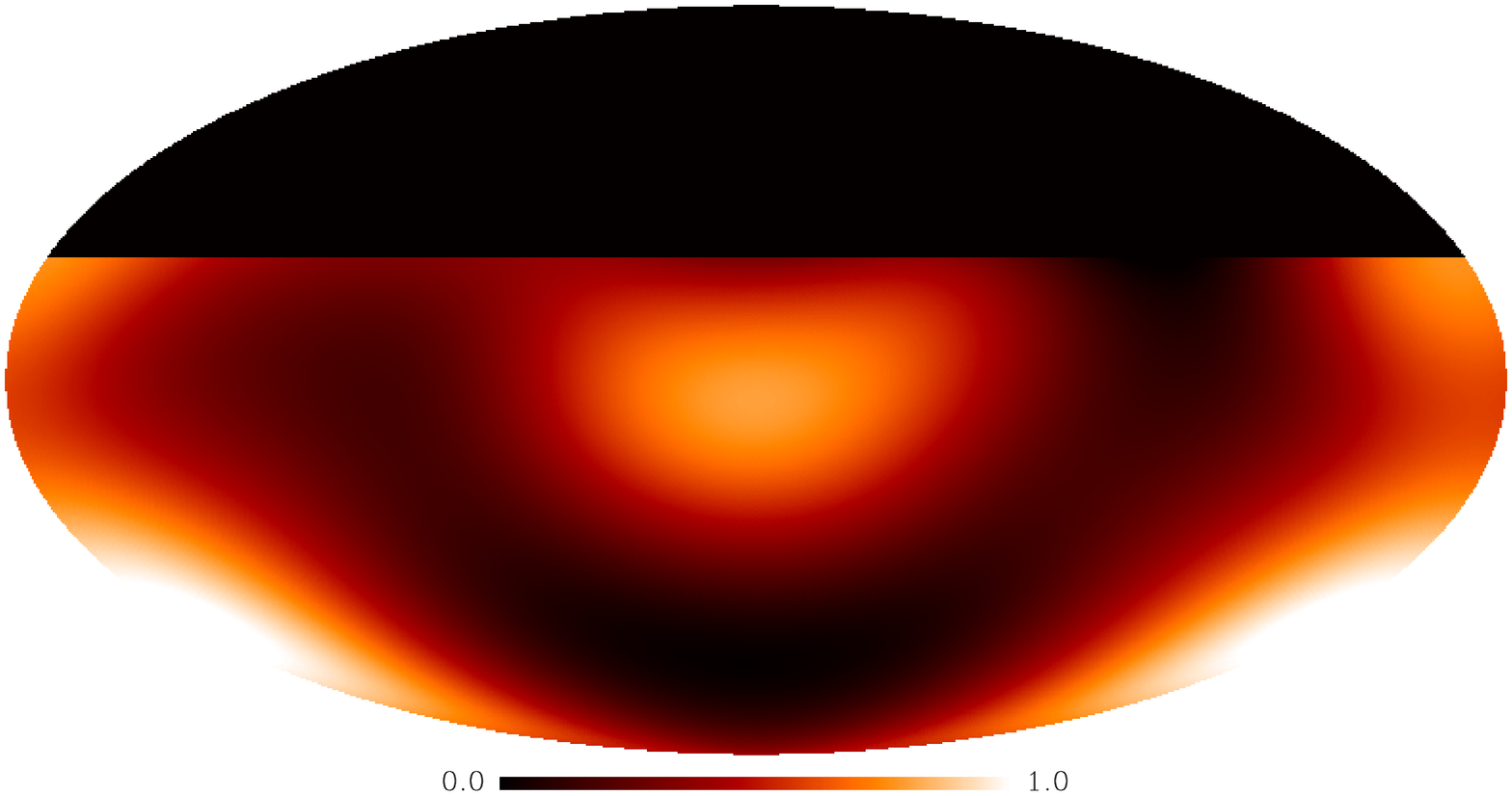}
	\includegraphics[width=4.5cm,angle=90]{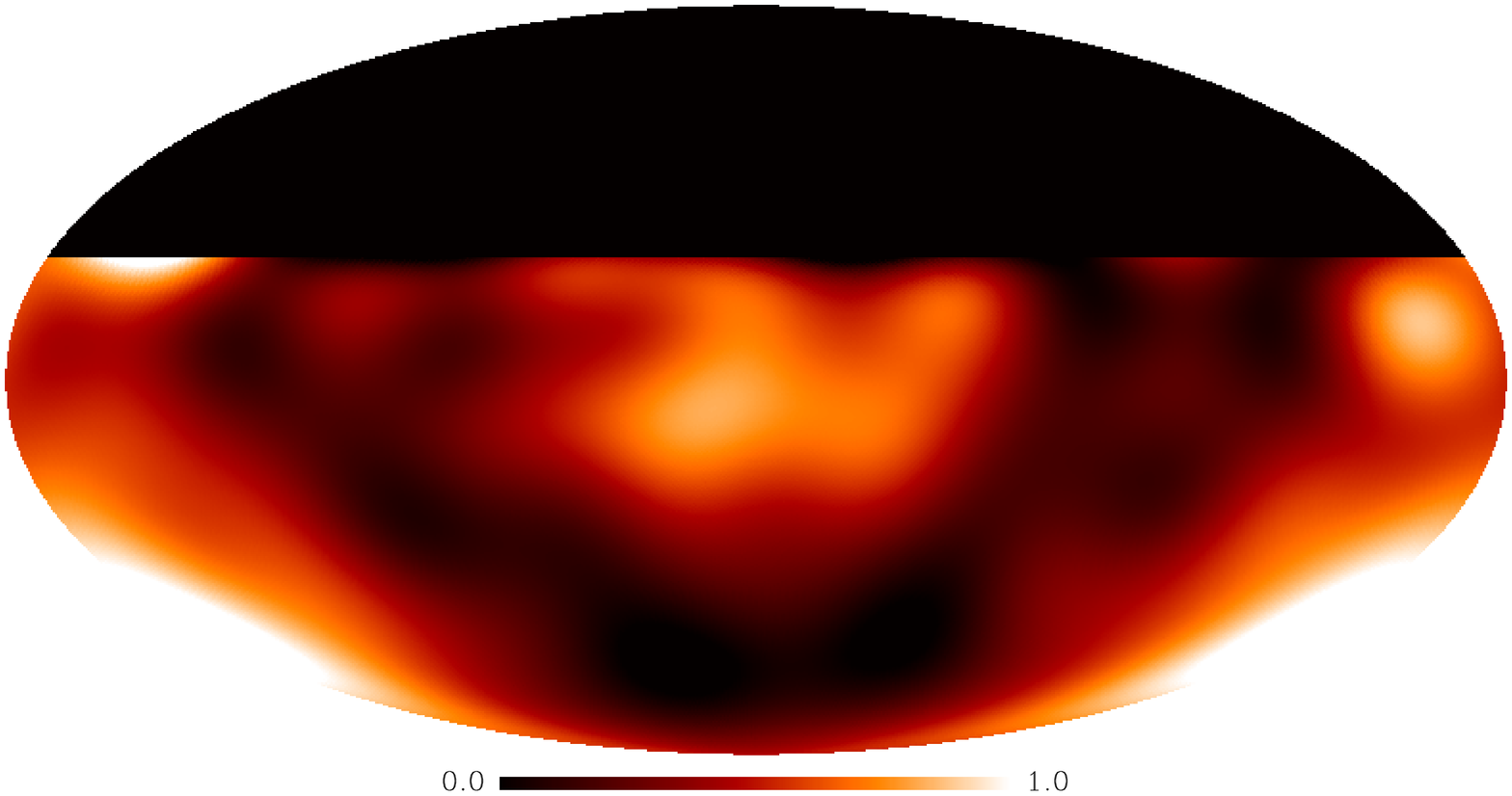}
	\caption{\small{\textit{Top left: Toy injected sky (combination of $Y_1^1$,$Y_2^2$ 
	and $Y_3^1$), in equatorial coordinates. Then recovered skies using different
	assumptions for $L$, and using the exposure function of Fig.\ref{fig:CovAugerSud}.
	Top right: Recovered sky assuming $L$=3. Bottom left: Recovered sky
	assuming $L$=5. Bottom right: Recovered sky assuming $L$=10. The unseen part of 
	the sky is hidden.}}}
	\label{fig:cover_region}
\end{figure}

We have shown that using a large value of $L$ in case of a partial coverage of
the sky forbids to give to any $a_{\ell m}$ coefficient an interpretation of
an individual multipolar moment. Nevertheless, one may wonder about the 
signification of the full set of coefficients $\{a_{\ell m}\}$. As a toy
example, we generated a distribution of points according to the exposure function
of Fig.\ref{fig:CovAugerSud} times the function shown on Fig.\ref{fig:cover_region} 
(top left) which is a combination of $Y_1^1$,$Y_2^2$ and $Y_3^1$. On top right of 
Fig.\ref{fig:cover_region}, we show the reconstructed sky assuming $L$ to be 
equals to 3, which illustrates that \emph{the reconstructed sky matches the 
injected one in the covered region even if the variance on each reconstructed
multipolar coefficient is already large (as shown in preceding sub-sections) for
$L$=3}. Increasing the value of $L$ to 5 (bottom left) or 10 (bottom right) do
not change this property of the expansion, as only additional statistical fluctuations
appear due to the finite number of points. On these plots, we hide the unseen part
of the sky, where the reconstructed expansion is meaningless. 

\subsection{Hypothesis test}

Any sky observed through an exposure function $\omega$ can thus be 
described precisely in the observed part of the sky by increasing L 
at a sufficient value. However the interpretation of each multipolar 
moment is problematic, because it depends strongly on the cut $L$. We want
now to build a statistical test to obtain a reasonable value of $L$ from the data themselves. 

Starting from an hypothesis on $L$ and the corresponding reconstructed
$\{a_{\ell m}\}$ coefficients, the likelihood function $\mathcal{L}_L$ 
built from the realization is \[
  \mathcal{L}_L=\prod_{i=1}^N \left( \omega(\Omega_i) \sum_{\ell m}^L
  \overline{a}_{\ell m} Y_\ell^m(\Omega_i) \right).
\]
For any particular realization, from this likelihood (which depends 
on $L$), we apply the method of the likelihood ratio to accept or to 
reject (within some chosen threshold) a null hypothesis 
$H_0 (\mathcal{L}_{L_0})$ with respect to another hypothesis
$H_1 (\mathcal{L}_{L_1})$ by computing \[
  \lambda = \frac{\mathcal{L}_{L_0}}{\mathcal{L}_{L_1}}.
\]
Asymptotically, for a sample obeying the hypothesis $H_0$, $-2 \ln{\lambda}$ is distributed according to a
$\chi^2$ with a number of degrees of freedom equal to the number
of extra parameters in the $H_1$ hypothesis with respect
to $H_0$. The value of $\lambda$ for any particular
realization can thus be used to validate (or to reject) an assumption
on $L$.

\begin{figure}[t]
  \centering
	\includegraphics[width=14cm,height=10cm]{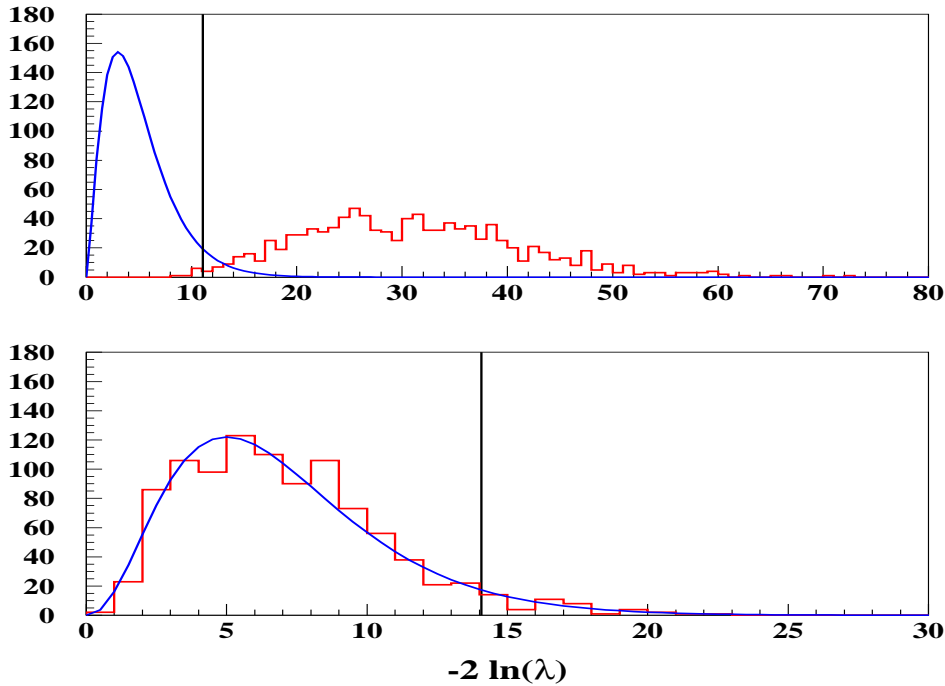}
	\caption{\small{\textit {Observed (red histograms) and expected 
	(continuous blue line) distributions for $\lambda$ in case of a quadrupolar injected
	pattern. Top: $L=2$ against the null hypothesis $L$=1.
        Bottom:  $L=3$ against the null hypothesis $L$=2. 
	The vertical line indicates the value of $\lambda$ below which
	the null hypothesis is accepted with a threshold at 5\%.}}}
	\label{fig:likeratio}
\end{figure}

As an example, let us assume that the cosmic rays follow a symmetrical quadrupolar
distribution $1+0.1\sin^2\theta -0.2\cos^2\theta$, and let us use
once again the exposure function shown on Fig.\ref{fig:CovAugerSud}. By 
restricting the reconstruction to a dipolar distribution, one then finds 
an artefact amplitude of about 5\%. To test the relevance of the hypothesis of a 
purely dipolar sky, one can thus - starting from this sky - estimate whether it 
is necessary or not to increase the degree of the expansion by calculating 
the ratio of the likelihood between the null hypothesis $L$=1 and another 
hypothesis on $L$, $L$=2 for instance. To show the behavior of the test, we 
generated 1000 different realizations of the quadrupolar pattern with
100,000 points each, then we reconstructed the parameters of the expansion 
within the two hypotheses, and finally computed the ratio of likelihoods. 
In this case, the hypothesis $L=2$ introduce 5 more parameters
$\{a_{2m}\}$, and the expected values of $-2\ln{\lambda}$ 
are asymptotically distributed as a $\chi^2$ with 5 degrees 
of freedom. We plot the result on the top of Fig.\ref{fig:likeratio}: 
by choosing the threshold of the test to be 5\% (vertical line at 
$-2\ln{\lambda}=11.07$), only 8 realizations over 1000 are accidentally 
accepted (red histogram). On the contrary, repeating the same procedure
to the $L$=2 null hypothesis with respect to the $L$=3 hypothesis, we 
show on the bottom of Fig.\ref{fig:likeratio} that the obtained
distribution perfectly matches the asymptotical expected one 
(a $\chi^2$ with 7 degrees of freedom in that case). With the same 
partial coverage, a similar test on samples of 1000
points gives a poor discrimination between different hypotheses, and
with only 100 points the test is completely irrelevant.

This procedure may be used to define a ``likely minimum value'' $L_{min}$ of $L$,
and to prevent a wrong interpretation of multipolar coefficients obtained with
a lower value, which are then biased (as the artefact dipole obtained above from
a symmetric quadrupole). Of course, a given sample cannot provide by itself an absolute
maximum for $L$, and in the presence of a hole the multipolar coefficients remain undefined
without an external assumption; however let us point out that in many cases, the values
of the coefficients at a given order have no intrinsic physical meaning if the distribution 
is of higher order. 

\section{Testing model predictions}

Let us consider a distribution $\lambda(\theta,\varphi)$ with coefficients
$a_{\ell m}$ on the $Y_{\ell}^m$; the observed distribution 
$\omega(\theta)\lambda(\theta,\varphi)$ has coefficients $\alpha_{\ell m}$ on the
$Y_{\ell}^m$, and the relation: \[
  a_{\ell m} = 
 \sum_{\ell'=\ell}^\infty\,C_{\ell^{\prime} \ell}^m\,\alpha_{\ell^{\prime} m} 
\]
may be inverted, because for each value of $m$ the matrix $C_{\ell \ell'}$ is
triangular, and the coefficients of the inverse relation may be computed exactly
for any value of $\ell$ : \[
  \alpha_{\ell m}=\sum_{\ell'=\ell}^\infty D_{\ell'\ell}^m\,a_{\ell' m} 
\]
The values of the $D_{\ell'\ell}^m$ are displayed in Fig.\ref{fig:coeff_llm}
(right) with the same example as on the left plot, but in linear scale: contrary
to the $C_{\ell'\ell}^m$, they remain below 1 in absolute value, and practically
negligible far away from the diagonal\footnote
{However, in practice, through the matrix inversion, the numerical divergence of the
$C_{\ell'\ell}^m$ limits the expansion to $L \simeq 15$ for this kind of coverage
function; this is sufficient for most studies on sky anisotropies.}.  

As a consequence, if a model gives predictions about the $a_{\ell m}$, it will be
possible, in some cases, to deduce predictions on the $\alpha_{\ell m}$,
which can be tested without any assumption on $L$. In that sense, the compatibility
of a model may be checked with observations over an incomplete sky with a precision
depending on the available statistics (but, of course, it can never prove that this
model is the only possible one).  

If a model makes a deterministic prediction, comparing the $\alpha_{\ell m}$ to
the predicted values may be a convenient way to test this model up to a given order of
multipolarity, that is, down to a given angular scale. The method is potentially more
interesting if the predictions are probabilistic. As we emphasized it in the 
introduction, this is a relevant framework to describe high energy cosmic rays 
physics. Indeed, even in a situation with a well-defined and structured configuration
of sources, propagation of cosmic rays unavoidably leads to a probabilistic 
nature of the obervable sky, that is to say, a probabilistic nature of the 
multipolar moments. Each class of models has intrinsically a natural variance 
encrypted in the $a_{\ell m}$ covariance matrix. Further, some models do not 
try to build a well-defined configuration of sources, but pick up randomly
cosmic rays at sources according to some distributions, making even more 
impossible to circumvent the characterization of a particular data set through 
a relevant statistical tool.

Consequently, the discrimination of models through an exploratory search in
a data set is potentially extremely powerful by looking for the distance 
of the full covariance matrix to the expected one. Most simple example is a 
model predicting random $a_{\ell m}$ following independent gaussian laws with 
variances $\sigma_{\ell}^2$. In that case, the covariance matrix for the 
$\alpha_{\ell m}$ reads \[
  \mathrm{cov}(\alpha_{\ell_1 m},\alpha_{\ell_2 m}) = \sum_{\ell'=m}^{\infty}
  D_{\ell'\ell_1}^m\, D_{\ell'\ell_2}^m~\sigma_{\ell'}^2  
\]
provided, of course, that this series converges: this is true if the series of
$\sigma_{\ell}^2$ converges (as it should do for physical models), and if the
$D_{\ell'\ell_1}^m$ have the behaviour suggested by Fig.\ref{fig:coeff_llm}.

\section{Conclusions}

To cope with a partial sky coverage, a formalism using the computation 
of moments on orthogonal functions was developed to recover the angular 
distribution of the incident flux from a sample of $N$ observed points. 
If the multipolar expansion is assumed to be upper bounded by $\ell\leq L$, 
the coefficients $a_{\ell m}$ may be estimated with a variance proportional 
to $1/N$ as usually, with a penalty factor increasing exponentially with 
$L$ if there is a hole in the coverage (but stabilizing rapidly if the coverage
is nowhere vanishing, even 
highly non-uniform). Two methods were tested, giving similar results, and 
practically the same variances.

Statistical tests based on likelihood ratios may be built to check an hypothesis
on the distribution, for example a given bound $\ell \leq L$.
 In any case, it is possible to express 
predictions of a model in terms of coefficients which can be computed without 
any assumption on $L$, and tested against the moments found with a sample of 
observed points.

The methods presented in this paper may be applied any cosmic ray dataset,
 provided that the arrival directions
and the coverage of the sky are known within a reasonable precision. 

\section*{Acknowledgments}

We thank members of the Auger collaboration for helpful discussions.
Fig.\ref{fig:cover_region} has been made using the HEALPix package~\cite{healpix}.

\end{document}